\documentclass[twocolumn,showpacs,aps,prl,
groupedaddress,amssymb,amsmath]{revtex4}

\usepackage{graphicx}
\usepackage{longtable}

\begin{document}

\title{Scaling behavior in Spiral Defect Chaos.}

\author{Kapilanjan Krishan}
\affiliation{Department of Physics and Astronomy, University of
California at Irvine, Irvine, California 92697-4575}

\date{9 March, 2007}

\begin{abstract}
We find the evolution toward power-law scaling in the distribution of roll lengths and nearest-neighbor distributions in a weakly turbulent regime of Rayleigh-Benard convection, known as spiral defect chaos. The state has a bounded domain of wave vectors in Fourier space attributed to the flow being highly confined to two dimensions. Our results indicates the existence of power-law scaling in the unconstrained horizontal dimension. The techniques described are broadly applicable to other pattern forming systems as well.
\end{abstract}

\maketitle

The patterns exhibited by spatially extended systems reflect the underlying dynamics. In non-equilibrium systems, complex patterns emerge at a macroscopic scale as a result of local nonlinearities~\cite{ch93}. Relating the dynamics from the microscopic scale of nonlinearities to the macroscopic patterns observed remains challenging. It is therefore important to develop generally applicable characterizations of a system that can capture the dynamics across a multiple scales in a robust and consistent way. 

The emergence of patterns represents the delineation of the system into regions with different physical characteristics. For instance, the domain structure of a material distinguishes between regions of differing orientation. The different regions may be considered as the different components comprising the pattern. In fluid systems, these structures are often in the form of variations in local velocities during the flow. For example, a characteristic of turbulent flows is the formation of eddies at a multitude of scales, in contrast, laminar flows have a well defined length scale associated with them~\cite{ll87,m90}. 

The flow of a fluid in highly constrained geometries is dominated by the confining boundaries. The reduction in size of a system often results in the suppression of turbulence through an effective reduction of the Reynolds number associated with the flow~\cite{m90}. While this reduction in size can be imposed along a single spatial dimension without affecting the other spatial dimensions, the influence on the resulting flow is not easily extracted.

In this letter, we extract power-law scalings associated with patterns representing fluid flow in Spiral Defect Chaos, a weakly turbulent state of Rayleigh-Benard convection~\cite{c89,mbcdg93,dpw94,mr07}. In this flow, the vertical dimension of the flow is highly constrained in comparison to the horizontal dimension. The convective instability is coincident with the formation of roll like structures with a wavelength on the order of the depth of the convective cell~\cite{c81,m90}. The system has received much attention as it shows unexpected chaotic behavior that is bistable with at values of system parameters corresponding to the flow being stable and stationary~\cite{as94,bca97,cepb97}. A number of characterizations of the state have been made emphasising local instabilities, wave-vector frustration and mean flows that may be linked with defect nucleation~\cite{cpc03,emb98,ct95,empe00,la96}.

We use two characterizations of the state that reveal length scales that are associated with the structure of the flow in spiral defect chaos. The important difference with earlier characterizations is that these measures reveal underlying power-law scalings in real space not identifiable with the distributions of the Fourier power spectrum. In the first characterization, we determine the scaling of the area of distinct convection roll structures. In the second characterization, we reduce the patterns exhibited by the system to a graph. This parameterizes the system through number of nearest-neighbors of each distinct roll.

\begin{figure}
\includegraphics[width=6cm]{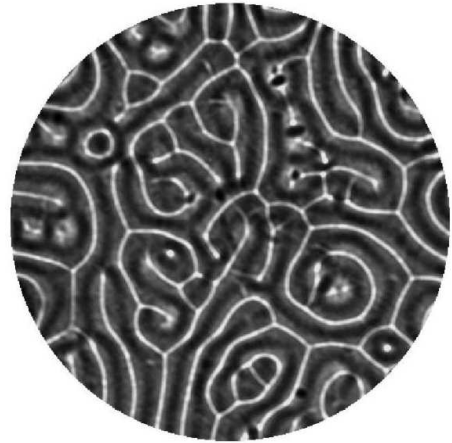}
\caption{A typical image of the convective state exhibiting spiral defect chaos(at $\epsilon=2.63$) as visualized using shadowgraphy. The convective fluid is bounded horizontally in a circular domain with radius $3.8{\rm cm}$ and vertically extends to about 0.69{\rm mm}} \label{sdc}
\end{figure}

Our experiments use compressed carbon di-oxide as the convective fluid. The gas is held in a convection cell bounded below by a 1cm thick gold plated aluminium mirror and above by a $2{\rm cm}$ thick transparent sapphire crystal. The lateral boundaries comprise of stacks of filter paper $0.069{\rm cm}$ thick that also act as a spacer between the upper and lower boundaries. The filter paper has a $3.8{\rm cm}$ diameter circular hole in the center within which the convection is visualized. The aluminium mirror has a resistive thin film heater attached to it on the bottom and rests on high precision screws that may be turned to align the mirror. The mirror is levelled with respect to the sapphire window using laser interferometry. Chilled water at a controlled temperature is circulated over the top of the sapphire window to maintain its temperature. In addition, thermistors embedded in the side of the aluminium mirror as well as close to the sapphire plate are used to monitor the temperature gradient across the convective layer. The setup in contained within an aluminium canister  pressurized to $30{\rm atm}$ with CO$_2$ gas using an external pressurized cylinder. Computer control of the thin film heater is used to set the temperature difference, $T$, between the top and bottom boundary of the convective cell to within $10{\rm mK}$ of the desired value. Our experimental setup is similar to that used earlier by Rogers {\it et. al.}~\cite{rsbs00,rsbp00}.

At a critical temperature difference across the fluid, $T = T_c$, the onset of convection occurs resulting in the formation of roll like structures. These rolls represent the transport of fluid vertically across the cell when thermal expansion induced buoyancy in the fluid overcomes dissipation due to viscosity and thermal conduction~\cite{c81,m90}. The visualization of the convective rolls is done using shadowgraphy~\cite{bbmthca96,s06}. This technique utilizes the variation in the refractive index between hot and cold regions of the gas to focus light differently. Light from a bright halogen lamp is passed through a pinhole and collimated using a concave lens. After passing through the convective layer, the spatial variation in light intensity is captured using a CCD array. The image is indicative of the changes in the refractive index of the light as it passes through the fluid. In our setup, bright and dark regions represent hotter upflows and cooler downflows.

We set our experiment at different values of $\epsilon (= (T-T_c)/T_c)$, and hold the temperature for over one hour to let transients dynamics die out. A sequence of 8000 images acquired at 11Hz are used in our data analysis. The total time the dynamics of the system is captured is many times the vertical thermal diffusion time (about 2 seconds in our system). The response of system falls qualitatively into three regimes~\cite{c89,p97}: low, $0 \le \epsilon \le 0.5$, where the convective rolls are stationary states of the system, intermediate values of $0.5 \le \epsilon \le 1.5$ result in the onset of time dependence and weak turbulence, and at $1.5 \le \epsilon$, the system shows prominent time-dependent behavior.

A background image of the fluid flow prior to the onset of convection is subtracted from every image to remove stationary optical inhomogeneities that may be associated with irregularities in the optical path(such as scratches in the mirror etc.). These difference images are Fourier filtered to eliminate the influence of high frequency noise associated with the CCD array as well as low frequency variations associated with nonuniform illumination. The intensity profile of the resultant image is representative of the structure of the convective rolls. The image is thresholded at the median value of intensity to yield a binary image. This choice of a threshold segregates the convective region into equal areas representing hot upflows and cold downflows to be consistent with the conservation of mass.

\begin{figure}
\includegraphics[width=4.5cm]{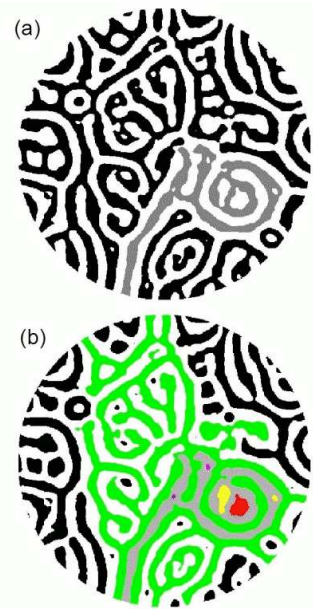}
\caption{In figure (a), a set of convection rolls representing the dark areas of figure~\ref{sdc} are indicated by dark regions of a binary image. As an example of a distinct component of this flow, one of the rolls that is singly connected is shaded in gray. In figure (b), the complementary flow is shown, representing the bright areas in figure~\ref{sdc}. In addition, the gray component is added for visual effect, while also indicating its nearest neighbors in color.} \label{binary_image}
\end{figure}

\begin{figure}[htb!]
\includegraphics[width=8cm]{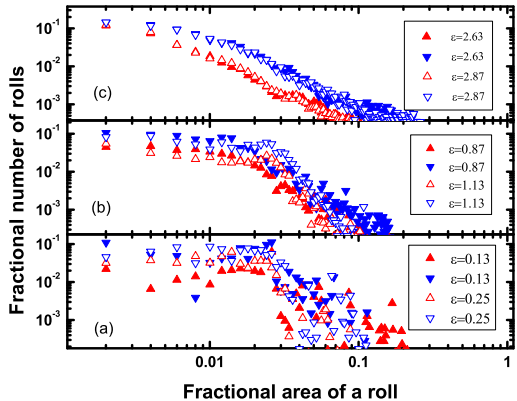}
\caption{The area of convective rolls scale as a power-law with the number of rolls. The slope in the plot (a) for both upflows and downflows is close to $-1.8$. For comparison, in plot(b), the curves are split into regions of slopes approximately $-0.35$ and $-3.6$ at $\epsilon = 0.87$. Upward pointing triangles refer to bright regions, and downward pointing triangles to dark regions of the binary images captured.} \label{kink}
\end{figure}

The segregation of the flow into two distinct components as depicted by the binary image(figure~\ref{binary_image}) provides the basis for our characterizations. As illustrated in figure~\ref{binary_image}a, each distinct bright(dark) roll is completely surrounded by dark(bright) rolls, or the boundary of the convective cell. In the first characterization, we simply look at the area of distinct bright or dark rolls as a fraction of the total convective area. A histogram of the number of distinct rolls at a given area fraction divided by the total number of distinct rolls vs their relative area at a given value of $\epsilon$ is plotted in figure~\ref{kink}. Since the convective rolls have a well defined average width, the area of the rolls is well approximated by a constant scalar multiple of the length of the roll. The convective rolls may be looked upon as primarily one dimensional curves within the circular convective region.

When $\epsilon \ge 1.5$, a power-law scaling of the roll areas to number of rolls emerges. However, prior to the occurrence of this scaling, we note two distinct scalings occur based on the area fraction of the rolls as seen in figure~\ref{kink}. The kink in these plots occur at an area corresponding to that of a roll with length equal to the diameter of the convective region. There is a sharp drop in the number of rolls that have a length greater than the diameter of the convection cell. This is because the rolls are without much curvature at lower values of $\epsilon$. At higher values of $\epsilon$, the roll curvature increases,  with an associated increase in the number of rolls with a larger area. The distinction between the distributions for upflows and downflows suggests that the flow is non-Boussinesq in nature~\cite{k05,g05,mr07}. The asymmetry between upflows and downflows increases with an increase in $\epsilon$. We find the system to be composed of a higher number of dark rather than bright rolls.

\begin{figure}[htb!]
\includegraphics[width=8cm]{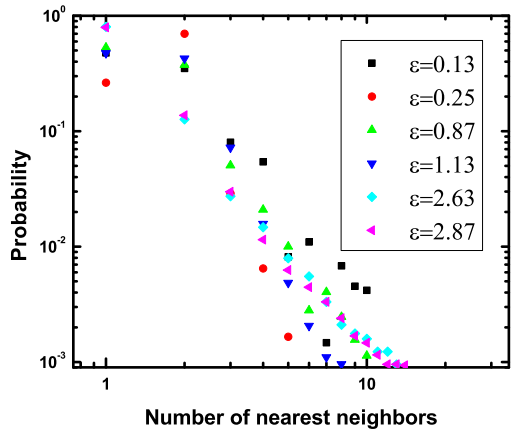}
\caption{The scaling of the number of nearest neighbors with increasing temperature gradient across the system. At high values of $\epsilon$, the system evolves to show a power-law scaling. The slope at the highest value of $\epsilon$ shown above is $-2.9$} \label{nn}
\end{figure}

A further topological characterization may be made by considering the binary image of the flow as a bipartite graph. In order to do this, we consider the number of distinct bright(dark) rolls each distinct dark(bright) roll shares an interface with. Figure~\ref{binary_image}b illustrates this measurement by indicating the nearest neighbors of a single convective roll in the flow(indicated in gray). In our binary image, bright(dark) regions are surrounded by dark(bright) regions with the only exceptions being at the boundaries. We count of the number of nearest neighbors every dark(bright) region has that is bright(dark). A set of numbers representing the number of nearest neighbors for each distinct dark/bright region is thus generated at a constant value of $\epsilon$. We normalize this set so their sum is unity to compute the probability of having a given number of nearest neighbors. This is ploted against the number of nearest neighbors for a convective roll as shown in figure~\ref{nn}. The probability of having a given number of neighbors scales as a power-law with the number of nearest neighbors. We see no significant difference between upflows and downflows in the scaling of the number of nearest-neighbors. This measurement is therefore insensitive to the breaking of the Boussinesq symmetry in contrast to the results presented in figure~\ref{kink}. Figure~\ref{nn} shows the scalings for bright regions of figure~\ref{binary_image}a and the results for the dark regions do not show significant differences.

The probability of having only one neighbor is finite for rolls that are at the boundary, or are completely encircled by a different roll. We note the general trend in figure~\ref{nn} for the number of rolls with two nearest-neighbors to decrease with increasing $\epsilon$. In the simple geometries of parallel rolls and concentric circles, the number of nearest-neighbors is two. In single spiral formations, the number  of nearest neighbors would be one, there being only a single bright roll and a single dark roll. When more complex patterns emerge, they can often be broken down into spatially localized regions with simple geometries~\cite{ehma95}. The present characterization develops a general technique to make global characterizations of the system based on the nearest-neighbor distributions within the flow. 

The experimental characterization we have used relies on a small image series at each value of $\epsilon$. It has been shown that the state of spiral defect chaos exhibits transitions between straight-roll states and curved roll states that are time dependent. This was done by observing the evolution of the system over a period of a month, keeping $\epsilon$ fixed~\cite{cepb97}. The importance of our study lies in extracting global measures that reveal an underlying scaling behavior for the various patterns exhibited. A more accurate parameterization of this scaling with $\epsilon$ would involve averaging over very large experimental data sets. Our current data suffices to demonstrate the existence of power-law behavior.

While the lengths of rolls show a sharp kink that is associated with the system size, there is no feature corresponding to this in the nearest-neighbor distributions. The scalings of the area of the system are metric dependent, and correspondingly show an associated length scale of the system size. The nearest-neighbor distributions however are metric independent and a purely topological measure. The two measures are related as they determine the packing of upflows and downflows in the system. A larger packing fraction of the rolls translates to an increase in the number of rolls. The increase in the number of distinct rolls with an increase in $\epsilon$ has been reported earlier~\cite{k05,g05}. 

The techniques presented allow for connecting dynamical phenomenon at various scales. Most models of fluid systems rely on the Navier-stokes equation, which describes the microscopic dynamics of the system. While the solutions to differential equations have been validated many times in comparison with experiments, it remains challenging to predict the large-scale structure of the fluid flow in a turbulent regime. Recently topological techniques have been used to model and characterize systems without reference to the microscopic dynamics. The present study develops characterizations that should offer insightful comparisons with other systems characterized using similar measures. In particular the evolution to power-law scaling has been of considerable interest~\cite{ws98,fff99,LotsOfPeople03}.

While developing numerical models, one is not always able to match the initial conditions or exact boundary influences that occur in experiments. The techniques outlined in this paper could also be used to quantify global statistical measures as a tool to compare experimental and numerical results. The measures described here provide average properties of the state of the system that complement other metrics currently in use~\cite{sc05}. Comparisons based on topological measures have the advantage of being scale-independent.

In spiral defect chaos, the extreme confinement of the fluid between the top and bottom plates primarily determines the wave-number exhibited by the system. This shows up as finite and bounded distribution of power in Fourier space, corresponding to the wavelength of the convective rolls. We find that while the wave-vectors are constrained, the length scales of the rolls and topological measures characterizing the nearest-neighbor distributions have power-law distributions.

The author would like to thank M. Gameiro, K. Mischaikow, N. Przulj and J. Scheel for useful discussions. In addition, the author is grateful to M. F. Schatz in whose laboratory the experiments were carried out.

\end{document}